\documentclass[aip, reprint,nofootinbib]{revtex4-2}

\usepackage{amsmath}  
\usepackage{amsfonts} 
\usepackage{graphicx} 
\usepackage{physics}  
\usepackage{cancel}
\usepackage{color}
\usepackage{xcolor}
\usepackage{multirow}
\usepackage{empheq}
\usepackage{mathtools}
\usepackage{upgreek}            
\usepackage{amssymb}
\usepackage{accents}
\usepackage[symbol]{footmisc}

\newcommand{\vect}[1]{\vb*{\va*{#1}}}   
\newcommand{\vectop}[1]{\vb*{\hat{\va*{#1}}}}   
\newcommand{\brb}[1]{\biggl({#1} \biggr)} 
\newcommand{\dverts}[1]{\lVert {#1} \rVert} 

\begin{document}


\title{From raw data to processed spectra: A step-by-step guide.}

\author{Erik F. Woering}
\affiliation{University of Groningen, Zernike Institute for Advanced Materials, Nijenborgh 3, 9747AG Groningen, The Netherlands}

\author{Richard Hildner}
\email{r.m.hildner@rug.nl}
\affiliation{University of Groningen, Zernike Institute for Advanced Materials, Nijenborgh 3, 9747AG Groningen, The Netherlands}


\date{\today}

\begin{abstract}
Optical spectroscopy is an important and widely used technique, for instance, to characterize new materials and to identify unknown compounds. Spectra are typically reported as a function of the wavelength of light, yet the information extracted from such spectra can be misleading. In contrast, spectra represented as a function of the frequency (or photon energy) allow for a more direct extraction of the intrinsic quantum-mechanical properties of the materials under investigation. Here we discuss this conversion for absorption, fluorescence and fluorescence excitation spectra. We show step-by-step the different factors that lead to a rescaling of the measured absorption and fluorescence signals.
\end{abstract}

\maketitle 


\section{Introduction}
Spectroscopy is a widely used technique to detect the response of a system to electromagnetic radiation as a function of the frequency (or wavelength).\cite{Lakowicz2006,Parson2015} For instance, the properties of core electrons of atoms can be probed by X-ray radiation, transitions of valence electrons of molecules are induced by radiation in the (near-)ultraviolet, visible and near-infrared (UV/Vis/NIR) range, molecular vibrations can be detected using infrared radiation, and nuclear spins are probed and manipulated by radio-frequency radiation. 

This places spectroscopy at an interesting position. On the one hand, spectroscopic techniques can be used to understand physics on the quantum scale; spectroscopic observations were the main drivers to develop quantum mechanics, allowing e.g. to interpret the spectral series of hydrogen by Lyman and Paschen.\cite{Lyman1904,Paschen1908} On the other hand, spectroscopy provides a means to identify and characterize new or unknown materials by their distinctive spectral response. Spectroscopy is used in a variety of fields, including physics, nanoscience, materials science, chemistry, and biology. It is therefore essential that the methodology and terminology are well-known. 

\begin{figure}[h!]
\centering
\includegraphics[width=8.5cm]{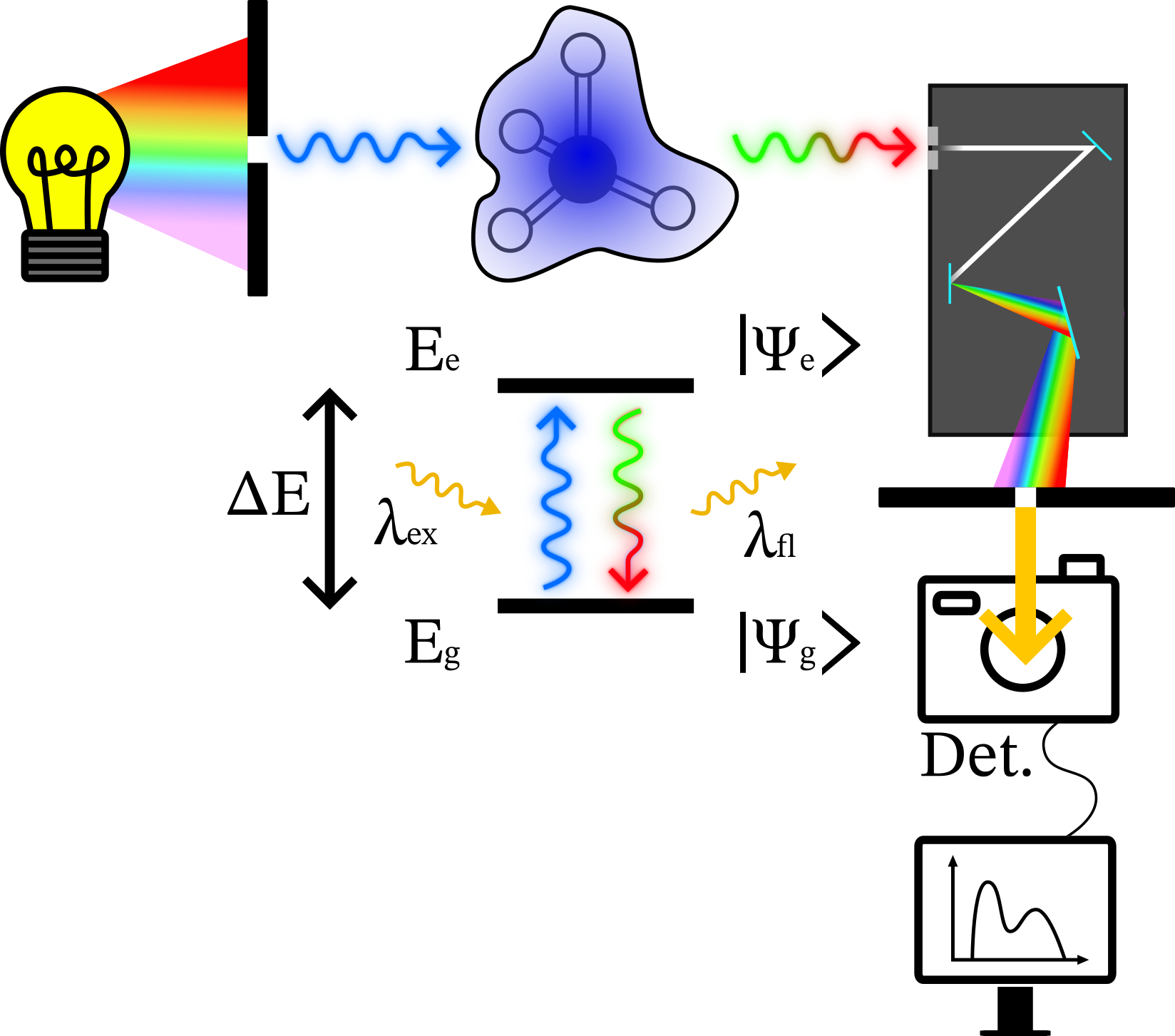}
\caption{(Color online) A schematic of a spectroscopic experiment. The output of a white-light source (`bulb') is filtered by a monochromator with a slit to select light of a specific wavelength interval (blue wavy arrow). This light induces transitions between energy levels of a molecule (for details, see text). The outgoing light (green/red wavy arrow) represents either transmitted light for absorption spectroscopy or emitted light for fluorescence spectroscopy. This light is recorded as a function of its wavelength using a combination of a spectrograph and a detector (Det.).}
\label{fig:Dipoles}
\end{figure}

In this paper, we focus on electronic spectroscopy, which refers to the interaction of electromagnetic radiation from the UV (150-400 nm), Vis (400-750 nm), and NIR (750-1000 nm) spectral ranges, or in short, light, with matter. As a concrete example, we will examine organic conjugated molecules in solution at room temperature, since it reflects the research of our group.\cite{Hildner2017} But the discussion is valid for other materials and for other spectral ranges too. Light incident on molecules in their electronic ground state can be absorbed and promote electronic transitions to excited states, if the photon energy $E_\mathrm{p}$ coincides with the energy difference $\Delta E$ between molecular energy levels, i.e., if the resonance condition $E_\mathrm{p} = \Delta E$ is satisfied (Fig.\,\ref{fig:Dipoles}). The inverse transition causes the emission of a photon (fluorescence) with a photon energy determined by the energy difference between the involved energy levels. $\Delta E$ is determined by the chemical structure of the molecule and can (in principle) be derived by quantum-mechanical calculations. The photon energy $E_\mathrm{p}$ is related to the frequency of light $\nu$, its wavelength $\lambda$, and wavenumber $\Tilde{\nu}$ by 
\begin{equation}
    E_\mathrm{p} = h \, \nu = \frac{h \, c_0}{\lambda} = h \, c_0 \, \Tilde{\nu},
    \label{eq:PhotonEnergy}
\end{equation}
with Planck's constant $h$ and the speed of light in vacuum $c_0$.\footnote{In the spectroscopy community wavelengths and wavenumbers are commonly given in units of nm and cm$^{-1}$, respectively, and photon energy is usually reported in eV, instead of the standard units m, m$^{-1}$, and J.} 

Spectra expressed as a function of wavelength are useful for deriving general trends. However, any quantitative discussion, regarding energy difference or linewidths for instance, should be done using spectra expressed as a function of energy, frequency or wavenumber. We will start with a short overview of the radiometric quantities that are measured in experiments, and we will briefly describe absorption, fluorescence and fluorescence excitation spectroscopic techniques (Section \ref{sec:Experimental}). Then we outline how to appropriately convert measured spectra from the wavelength to the energy scale, and we show how to connect spectroscopic measurements to intrinsic quantum-mechanical properties of molecules  (Section \ref{sec:Quantum}). Finally, we illustrate those conversions using a worked, step-by-step example (Section \ref{sec:Relevance}). An important point in all sections is that we emphasise the use of units from the \textit{International System of Units}. We will point out differences to more commonly used units where appropriate, e.g. the cgs system for electromagnetic fields and the `spectroscopic units' for the Beer-Lambert law. Although we do not report new findings here, we believe that a concise presentation of this topic is missing to date, because this information is typically spread across several textbooks and papers.\cite{Lakowicz2006,Parson2015,Hilborn1982,McCumber1964,Ryu2024,Fowler1962} We hope that this article will help instructors to clearly present the best methodology for converting and interpreting spectra. 

\section{Experimental Approaches}
\label{sec:Experimental}
\subsection{What do we measure: Important spectroscopic quantities}
\label{sec:Measure}

Before delving into the details of spectroscopic techniques and conversion factors, it is instructive to briefly recap quantities that characterize electromagnetic radiation. Strictly speaking, only radiometric quantities that are absolute measures of electromagnetic radiation and that are defined in the \textit{International System of Units}, should be used.\footnote{Photometric units are measures of electromagnetic radiation that are ``scaled'' by the wavelength-dependent sensitivity of a ``standard'' human eye. But those units are usually not relevant in the context of quantitative spectroscopy.} A basic quantity is the radiant energy $Q_\mathrm{e}(\nu)$ (in J), which is the energy of the electromagnetic radiation and which can be related to the photon energy via $Q_\mathrm{e} = N\,E_\mathrm{p}$, where $N$ is the total number of photons. The radiant energy density $\rho_\mathrm{e}(\nu)$ (in J\,m$^{-3}$) is the radiant energy per unit volume. Another important quantity is the radiant flux or radiant power $P_\mathrm{e}(\nu)$ (in W), which corresponds to the radiant energy emitted, reflected, transmitted or received per unit time. The irradiance $\mathcal{E}_\mathrm{e}(\nu)$ (in W\,m$^{-2}$) is the radiant power received by a surface of unit area. The subscript ``$\mathrm{e}$'' denotes radiometric quantities (``$\mathrm{e}$'' stands for ``energetic''), while the subscript ``$\mathrm{p}$'', see e.g. Eq.\,(\ref{eq:PhotonEnergy}), indicates quantities related to photons. The word ``spectral'' and the subscript $\nu$ (or $\lambda$) is added (e.g. spectral irradiance $\mathcal{E}_{\mathrm{e},\nu}$) to indicate that the quantity is per unit frequency (or per unit wavelength, depending on the spectral unit used). Most quantities have directional counterparts, i.e., per unit solid angle (in sr) along a specific direction, customarily denoted by the subscript $\Omega$. The directional counterpart of the radiant power is the radiant intensity $I_{\mathrm{e},\Omega}$ (in W\,sr$^{-1}$), i.e., the radiant power leaving a source along a given direction within an elementary solid angle $d\Omega$. For the irradiance, the corresponding directional quantity is the radiance $L_{\mathrm{e},\Omega}$ (in W\,m\textsuperscript{-2} sr\textsuperscript{-1}). We note that there sometimes is confusion in terminology, as especially irradiance is often referred to as ``intensity''. For a more complete list of radiometric quantities, we refer the interested reader to the online platform of the International Organization of Standardisation (ISO),\cite{ISO} or publications from IUPAC,\cite{McNaught2019} NIST,\cite{Newell2019} IEC,\cite{IEC2020} or CIE.\cite{Thorseth2023}

\begin{figure}[h!]
\centering
\includegraphics[width=8.5cm]{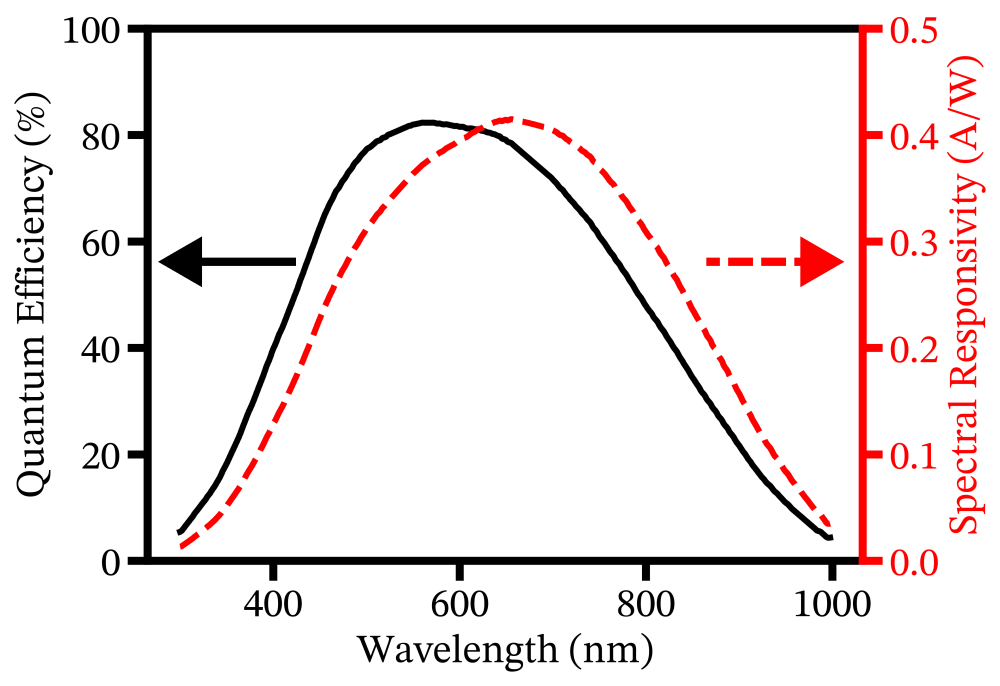}
\caption{(Color online) Quantum efficiency (black, solid) and spectral responsivity (red, dashed) for a scientific CMOS camera. Data taken from Ref.\cite{Zyla}}
\label{fig:EQVersusSR}
\end{figure}

Commonly used detectors, such as photomultiplier tubes (PMT), charge-coupled device (CCD) arrays, or CMOS detectors, convert incoming photons into electrons and then into an electrical signal (voltage or current), often involving an amplification step.\cite{Moore2009} The number of generated electrons is proportional to the number of photons received by the detector's active area. The proportionality factor, the so-called quantum efficiency $\eta_\mathrm{Q}(\lambda)$, is a (strong) function of the wavelength (most spectrometers output wavelength-dependent signals, hence such factors are often only provided as a function of $\lambda$ by manufacturers). As an example, the black solid curve in Fig. \ref{fig:EQVersusSR} shows the wavelength-dependent quantum efficiency in the UV/Vis/NIR range for a scientific CMOS camera (Zyla 4.2 Plus sCMOS, Andor). From the produced electron flux, i.e., the number of electrons output per unit time, $q_{\mathrm{el}}(\lambda)$, we obtain the photon flux, $q_\mathrm{p}(\lambda)$, received by the detector via the quantum efficiency, $q_\mathrm{p}(\lambda) = q_{\mathrm{el}}(\lambda) \, (100 \%)/\eta_\mathrm{Q}(\lambda)$.\footnote{Mathematically, $\eta_Q$ assumes a value smaller or equal to $1$, but for convenience it is often provided in percent.}

Instead of the quantum efficiency, manufacturers often provide the wavelength-dependent spectral responsivity $\mathcal{R}(\lambda)$ of the detector (in units A\,W$^{-1}$).\footnote{For thermopile sensors, or thermal power sensors, the responsivity is given in V\,W$^{-1}$, since a voltage is generated by absorbed light in a thermocouple. Let us also note that the spectral responsivity is sometimes improperly referred to as ``sensitivity''. This latter term holds a different meaning: The sensitivity of a device is the lowest input signal needed to be able to produce an output larger than noise.} This factor relates the electrical current $\mathcal{I}(\lambda)$ output in response to the radiant power received via $P_\mathrm{e}(\lambda) = \mathcal{I}(\lambda) / \mathcal{R}(\lambda)$. Fig.\,\ref{fig:EQVersusSR} shows the wavelength-dependent responsivity for the scientific CMOS camera (red dashed line). The quantum efficiency and the spectral responsivity are related via
\begin{equation}
  \eta_\mathrm{Q}(\lambda) = \mathcal{R}(\lambda) \, \frac{h \, c_0}{\lambda \, e}, 
  \label{eq:QE_responsivity}
\end{equation}
with $e$ (in C) being the elementary charge. 

In passing we note that the correction by the wavelength-dependent quantum efficiency or spectral responsivity is usually done automatically in commercial spectrometers and, for instance, the detected number of photons or the detected radiant power (or irradiance) as a function of the wavelength is displayed (though the latter often termed ``intensity'' without specifying whether power or irradiance is meant). For some detectors the correction can be switched via their control software between displaying photon counts and power. Nevertheless, it is very useful and instructive (and recommended!) to carefully check for each device which quantity is detected and displayed. For this check, spectra of well known reference compounds, such as laser dyes, can be measured and compared to data in literature.\cite{Taniguchi2017, Brackmann2000} For home-built setups this (efficiency, responsivity) correction must be done as first step in data analysis by the experimenter.


\subsection{How do we measure: Spectroscopic techniques}
\label{sec:SpecTec}
In the following, we briefly describe three of the most important spectroscopic techniques, absorption, fluorescence and fluorescence excitation spectroscopy, and detail the (wavelength-dependent) quantities measured in those experiments. 

\begin{figure}[h!]
\centering
\includegraphics[width=8.5cm]{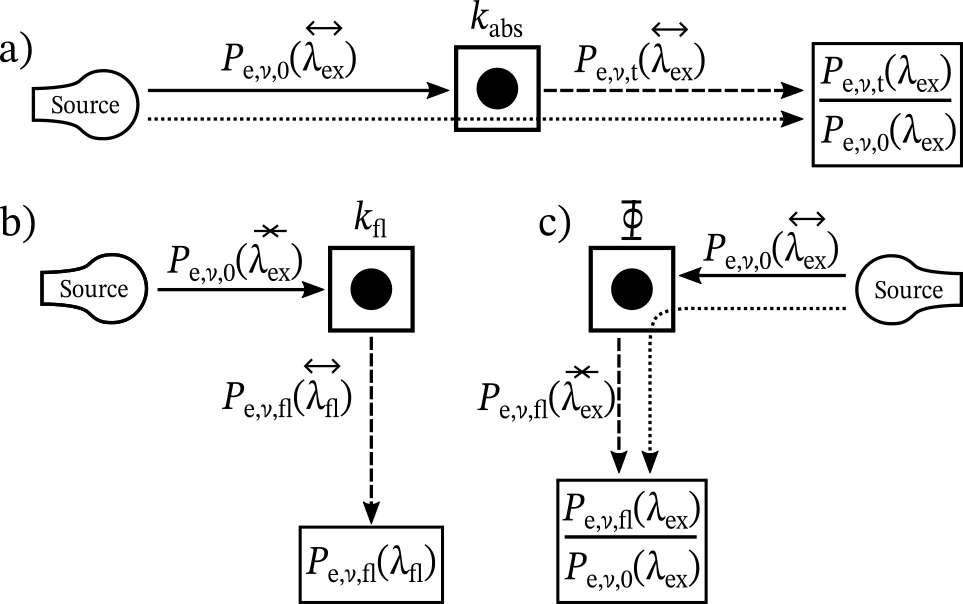}
\caption{Schematics of spectroscopic techniques. a) Absorption spectroscopy. b) Fluorescence spectroscopy. c) Fluorescence excitation spectroscopy. Light paths are indicated with solid arrows for incident light, dashed arrows for detected light, and dotted arrows for light from a reference measurement. The double-headed arrows above $\lambda$ indicate that it is a function of the (excitation or fluorescence) wavelength; the crossed line indicates that the signal is taken at a fixed wavelength (interval). For further details see text.}
\label{fig:ExcitationSpectrum}
\end{figure}

Using \textbf{absorption spectroscopy} (also known as UV/Vis/NIR or electronic spectroscopy) one measures which light wavelengths are absorbed by molecules. To record the full absorption spectrum, a broadband light source is used with a combination of a grating and a slit to select a wavelength interval $\dd\lambda_{\mathrm{ex}}$ around a central wavelength $\lambda_{\mathrm{ex}}$ which will excite the sample. This excitation wavelength $\lambda_{\mathrm{ex}}$ is then scanned across the relevant UV/Vis/NIR range and, for instance, the transmitted spectral radiant power $P_{\mathrm{e},\nu,\mathrm{t}}(\lambda_{\mathrm{ex}})$ is recorded after passing through the sample. However, the spectral radiant power incident on the sample is often not identical for all wavelengths, but it is determined by the (strongly) wavelength-dependent output of the light sources used. Moreover, the sample container (e.g.\,a cuvette) or the solvent in which the absorbing molecules of interest are dissolved can absorb light. Due to the absorption spectra of the common cuvette materials and of the usual solvents, this is especially critical in the UV range. To correct for those factors, the ratio is taken between the transmitted spectral radiant power $P_{\mathrm{e},\nu,\mathrm{t}}(\lambda_{\mathrm{ex}})$ and the reference spectral radiant power $P_{\mathrm{e},\nu,0}(\lambda_{\mathrm{ex}})$. The latter is measured after the light traverses a reference path that includes a container filled with solvent but no absorbing molecules (Fig.\,\ref{fig:ExcitationSpectrum}a). Typically, either the transmittance $T(\lambda_{\mathrm{ex}})$ or the absorbance\footnote{Besides ``absorbance'', the term ``optical density'' is still commonly used in the spectroscopy community, yet IUPAC discourages its use.} $A(\lambda_{\mathrm{ex}})$ is displayed:\footnote{Traditionally transmittance is more common in IR spectroscopy, whereas for UV/VIS/NIR spectroscopy it is absorbance.}
\begin{eqnarray}\label{eq:TandA}
T(\lambda_{\mathrm{ex}}) &=& \frac{P_{\mathrm{e},\nu,\mathrm{t}}(\lambda_{\mathrm{ex}})}{P_{\mathrm{e},\nu,0}(\lambda_{\mathrm{ex}})} \quad \text{and} \nonumber\\ A(\lambda_{\mathrm{ex}}) &=& \log_{10}\left(\frac{P_{\mathrm{e},\nu,0}(\lambda_{\mathrm{ex}})}{P_{\mathrm{e},\nu,\mathrm{t}}(\lambda_{\mathrm{ex}})}\right) = -\log_{10}T(\lambda_{\mathrm{ex}}).
\end{eqnarray}
Both quantities are dimensionless.\footnote{Please take note of the difference between dimensionless units, i.e., all units cancel out in dimensional analysis, and arbitrary units, i.e., non-standard units used for convenience, for instance when normalizing data.} Note that due to this ratiometric measurement, it is irrelevant which radiometric quantity we use here, since all are related by constant factors in a given experiment (see the first paragraph in section \ref{sec:Measure}).

The absorbance is related to the wavelength-dependent molar absorption coefficient $\varepsilon(\lambda_{\mathrm{ex}})$ (in m$^2$\,mol$^{-1}$), the concentration of absorbing molecules $\mathcal{C}$ (in mol\,m$^{-3}$), and the path length of the sample container $l$ (in m) via the Beer-Lambert-law\cite{Parson2015} 
\begin{equation}
  A(\lambda_{\mathrm{ex}}) = \varepsilon(\lambda_{\mathrm{ex}}) \, \mathcal{C} \, l. 
  \label{eq:Beer-Lambert}
\end{equation}
The molar absorption coefficient is a characteristic quantity of a molecule, associated with a transition between specific energy levels. 

In \textbf{fluorescence spectroscopy} (also know as fluorescence emission spectroscopy or emission spectroscopy) the light spontaneously emitted by molecules via radiative decay into the ground state is detected. Molecules are excited with a fixed excitation wavelength $\lambda_{\mathrm{ex}}$ to induce a transition to an excited state. According to Kasha's rule, fluorescence originates from the transition from the lowest-energy excited state to the ground state.\cite{Kasha1950} The fluorescence is collected under an angle of 90$^\circ$ with respect to the direction of the excitation light, which avoids direct propagation of the excitation light to the detector.  To record the fluorescence spectrum, the fluorescence signal $S(\lambda_{\mathrm{fl}})$ is detected as a function of the fluorescence wavelength $\lambda_{\mathrm{fl}}$ (Fig. \ref{fig:ExcitationSpectrum}b) by dispersing it in a monochromator and detecting in each wavelength interval $\dd\lambda_{\mathrm{fl}}$ after a slit (Fig.\,\ref{fig:Dipoles}). For instance, a spectral radiant power as a function of the wavelength is then obtained, $S(\lambda_{\mathrm{fl}}) = P_{\mathrm{e},\nu,\mathrm{fl}}(\lambda_{\mathrm{fl}})$.

In fluorescence spectroscopy, it is important to consider instrumental effects prior to data analysis: The gratings used in monochromators possess wavelength-dependent diffraction efficiencies $\eta_\mathrm{D}(\lambda)$ that depend on the groove density, the blaze angle (blaze wavelength), and the polarisation of light. If other optical elements, like optical filters with a wavelength-dependent transmittance $T_\mathrm{O}(\lambda)$ or mirrors with reflectance $R_\mathrm{O}(\lambda)$ are present, they must be considered, too. With this in mind, a correction function can be defined as 
\begin{equation}
\Theta(\lambda_{\mathrm{fl}}) = \brb{\eta_\mathrm{D}(\lambda_{\mathrm{fl}}) \, T_\mathrm{O}(\lambda_{\mathrm{fl}}) \, R_\mathrm{O}(\lambda_{\mathrm{fl}})}^{-1}.
\end{equation}
To obtain the signal $S(\lambda_{\mathrm{fl}})$ that represents the fluorescence spectrum free of instrumental effects, the detected (raw) signal $S_\mathrm{d}(\lambda_{\mathrm{fl}})$ must be corrected by applying this function. Finally, the detector's quantum efficiency or spectral responsivity [see Eq.\,(\ref{eq:QE_responsivity})] should also be taken into account:
\begin{equation}\label{eq:CorrectedSpectrum}
    S(\lambda_{\mathrm{fl}})=
    \begin{cases}
    \left(\frac{\Theta(\lambda_{\mathrm{fl}}) }{ \eta_{Q}(\lambda_{\mathrm{fl}}) } \right) \, S_\mathrm{d}(\lambda_{\mathrm{fl}}) & \text{if $S_\mathrm{d}$ is a photon flux},\\
     \left(\frac{\Theta(\lambda_{\mathrm{fl}}) }{ \mathcal{R}(\lambda_{\mathrm{fl}}) } \right) \, S_\mathrm{d}(\lambda_{\mathrm{fl}})& \text{if $S_\mathrm{d}$ is a radiant power.}
    \end{cases}
\end{equation}
Whereas in commercial spectrometers such instrumental factors are automatically corrected via a stored correction function, for home-built setups they must be considered explicitly.

\textbf{Fluorescence excitation spectroscopy} (also known as fluorescence-detected absorption spectroscopy) is used to obtain the absorption spectrum from highly dilute samples, for which the absorbance $A(\lambda_{\mathrm{ex}})$ is very small and cannot be detected by standard absorption measurements. This approach is regularly used in the field of single-molecule spectroscopy.\cite{Orrit1990} Here, the molecules are excited at an excitation wavelength $\lambda_{\mathrm{ex}}$ (with a bandwidth of $\dd\lambda_{\mathrm{ex}}$) that is scanned, and the fluorescence signal $S_{\mathrm{fl}}(\lambda_{\mathrm{ex}})$ is recorded as a function of the excitation wavelength (Fig.\,\ref{fig:ExcitationSpectrum}c). $S_{\mathrm{fl}}(\lambda_{\mathrm{ex}})$ is detected either at a specific fixed fluorescence wavelength interval $\dd \lambda_{\mathrm{fl}}$, or within a broader, fixed wavelength range. For instance, $S_{\mathrm{fl}}(\lambda_{\mathrm{ex}})$ can be the spectral radiant power $P_{\mathrm{e},\nu,\mathrm{fl}}(\lambda_{\mathrm{ex}})$. This approach exploits the fact that the detected fluorescence signal is proportional to the photon flux, see Eq.\,(\ref{eq:CorrectedSpectrum}), which, in turn, is proportional to the number of fluorescence photons emitted per unit time (per molecule), $k_{\mathrm{fl}}$. Via the fluorescence quantum yield of the molecules $\Phi = \frac{k_{\mathrm{fl}}}{k_{\mathrm{abs}}}$,\cite{Lakowicz2006,Parson2015} the detected fluorescence signal is then proportional to the number of absorbed photons per unit time (per molecule), $k_{\mathrm{abs}}$.\footnote{We consider here only molecules for which the fluorescence quantum yield $\Phi$ is independent of the excitation wavelength $\lambda_{\mathrm{ex}}$. For complex systems, in which, for instance, energy transfer between (several) fluorescent entities takes place, $\Phi$ can vary and render the analysis of excitation spectra rather difficult.} As we will show further below, the absorption rate $k_{\mathrm{abs}}$ at each $\lambda_{\mathrm{ex}}$ is directly related to the molar absorption coefficient $\varepsilon(\lambda_{\mathrm{ex}})$ and thus to the absorbance $A(\lambda_{\mathrm{ex}})$, hence the term fluorescence-detected absorption. Note that it is important to normalise the measured fluorescence signal by the incident spectral radiant power $P_{\mathrm{e},\nu,0}(\lambda_{\mathrm{ex}})$ of the light source at each excitation wavelength, as this power is typically not constant over the relevant spectral range.

Finally, we note that further subtleties of spectroscopic measurements, such as primary and secondary inner filter effects, issues related to fluorescence quantum yields, absorption measurements at very high/very low concentrations and excitation spectroscopy at high concentrations, are discussed in detail in excellent tutorial reviews and textbooks.\cite{Levitus2020, Lakowicz2006}


\subsection{Wavelength-to-frequency conversion: Jacobian transformation}
\label{sec:Jac}
Having recorded a spectrum as a function of the (excitation or fluorescence) wavelength $\lambda$, the first step before further quantitative analysis is the conversion into a representation as a function of the frequency $\nu$ or photon energy $E_\mathrm{p}$ (or any other quantity directly proportional to frequency). Since one measures spectral radiometric quantities, with signals being acquired within a certain wavelength interval $\dd\lambda$ around $\lambda$, one obtains spectra sampled at discrete wavelengths. Energy conservation dictates that the total signal as a function of $\lambda$, i.e., the signal integrated over the experimental wavelength range, must be identical to the signal integrated over the corresponding frequency range. Hence, we write:
\begin{eqnarray}\label{eq:AreaConservation}
\int S(\nu) \dd\nu &=& \abs{\int S(\lambda)\dd\lambda}, \quad \text{or equivalently,} \nonumber\\ 
S(\nu) \dd \nu &=& \abs{S(\lambda) \dd\lambda}.
\end{eqnarray}
The absolute magnitude avoids the sign change due to the opposite integration directions in wavelength and frequency space. Using Eq.\,(\ref{eq:PhotonEnergy}), this yields
\begin{eqnarray}
    S(\nu) = \abs{S(\lambda) \dv{\lambda}{\nu}} &=& \abs{S(c_0\nu^{-1}) \dv{}{\nu} \left (c_0\nu^{-1}\right )}\nonumber\\ &=& \abs{S(c_0\nu^{-1}) \frac{c_0}{\nu^{2}}}.
    \label{eq:JacobianConversion}
\end{eqnarray}
Eq.\,(\ref{eq:JacobianConversion}) is known as the Jacobian transformation.\cite{Mooney2013} In the second step the substitution of the signal's argument, $\lambda \rightarrow c_0 \, \nu^{-1}$, indicates that the wavelength axis of the measured signal is replaced by the corresponding frequency axis, while the signal (the y-axis) at each wavelength (frequency) interval does not change (see also below section \ref{sec:Relevance}). 

For the techniques described in section \ref{sec:SpecTec}, this implies that for absorption spectra the conversion factors in Eq.\,(\ref{eq:JacobianConversion}) cancel out, since the absorbance (or transmittance) represents the ratio of two signals taken at identical excitation wavelengths. Hence, only the x-axis is to be converted from $\lambda$ to $\nu$ (or $E_\mathrm{p}$). For fluorescence spectra, the Jacobian transformation is highly relevant, however, and will change the shape of the spectra due to the $\nu^{-2}$ scaling in Eq.\,(\ref{eq:JacobianConversion}). For fluorescence excitation spectra, this transformation is only relevant for the spectral radiant power of the excitation source, the wavelength of which is scanned; the fluorescence signal as function of the excitation wavelength is taken at a constant fluorescence wavelength (interval), which thus does not require a transformation.


\section{Optical Transitions: Connecting Quantum Mechanics with Measured Quantities}
\label{sec:Quantum}
\subsection{Absorption}
\label{sec:Abs}
For a simplified description of absorption processes, we exploit the common semi-classical approach, which models the molecule quantum mechanically, by a time-independent Hamiltonian $\hat{H}_0$, and describes the light field classically.\cite{Sakurai2020} The molecule is approximated as a two-level system with a ground state and an excited state, with the corresponding set of (stationary) wavefunctions $\{\ket{\Psi_\mathrm{g}}, \ket{\Psi_\mathrm{e}}\}$, and eigenenergies $E_\mathrm{g}$ and $E_\mathrm{e}$ for the ground state $\mathrm{g}$ and the excited state $\mathrm{e}$, respectively (Fig.\,\ref{fig:Dipoles}). Transitions from the ground state to the excited state are induced by interaction with light at a frequency $\nu$, for which the electric field vector reads $\vect{E}(t) = 2 \, \vect{E}\mathbf{_0} \, \cos{(2 \, \pi \, \nu \, t)},$ with $\vect{E}_\mathbf{0}$ being the peak amplitude of the field. Note that we drop the subscript ``$\mathrm{ex}$'' for denoting excitation frequencies (wavelengths) from here on, for simplicity. The light field must be weak in the sense that it does not modify the eigenenergies and wavefunctions of $\hat{H}_0$, but it induces transitions if the resonance condition $E_\mathrm{e} - E_\mathrm{g} = h \, \nu$ is satisfied. The interaction between light and molecule is described by the time-dependent interaction Hamiltonian $\hat{H}'(t)$, derived from the classical interaction energy between an electric dipole and an electric field
\begin{equation}
    \hat{H}'(t) = -\vect{E}(t) \cdot \vectop{\mu}.
    \label{eq:DipoleOperator}
\end{equation}
Here, $\vectop{\mu}$ is the dipole operator that reflects the charge distribution of electrons and nuclei in the molecule. 

To obtain a connection between the intrinsic quantum mechanical properties of the molecule, the absorption rate $k_{\mathrm{abs}}(\nu)$ and the spectral shape of the absorption peak, we have to evaluate the time-dependent Schr\"odinger equation using the full time-dependent Hamiltonian for our system, $\hat{H}(t) = \hat{H_0} + \hat{H}'(t)$. This evaluation yields (see supplementary material section A for full details):
\begin{equation}
    D(\nu) = \frac{3 \, \varepsilon_0 \, c_0 \, h^{2}}{2\pi^2} \, \frac{n(\nu)}{\mathcal{E}_{\mathrm{e},\nu}(\nu)} \, k_{\mathrm{abs}}(\nu).
    \label{eq:UprateIrradiance}
\end{equation}
Here, $D(\nu)$ is the transition dipole strength defined as
\begin{equation}
    D(\nu) \equiv \dverts{ \bra{\psi_{\mathrm{e}}} \; \vectop{\mu} \; \ket{\psi_{\mathrm{g}}}}^{2} = \dverts{\vect{\mu}}^{2},
    \label{eq:DipoleStrength}
\end{equation}
with the transition dipole moment $\vect{\mu}$, which is a fundamental quantum-mechanical property of a molecule. $D(\nu)$ describes the probability of absorption at frequency $\nu$,\cite{McNaught2019} i.e., how ``efficiently'' a molecule interacts with the light field. $n(\nu)$ is the (dimensionless) refractive index of the medium (solvent) surrounding the molecules, $\mathcal{E}_{\mathrm{e},\nu}(\nu)$ is the spectral irradiance of the light incident on the molecule, and $\varepsilon_0$ is the vacuum permittivity.

Unfortunately, Eq.\,(\ref{eq:UprateIrradiance}) is still of no practical use, since we cannot directly measure the absorption rate. Moreover, in a real sample with many absorbing molecules, the spectral irradiance $\mathcal{E}_{\mathrm{e},\nu}(\nu)$ continuously decreases with propagation distance due to continuous absorption of photons.\footnote{We assume here for simplicity that the incident irradiance is identical to the reference irradiance, see Fig.\,\ref{fig:ExcitationSpectrum}a. If the absorption of the cuvette and/or the solvent becomes important, this can be easily included here.} This decrease can be described by the Beer-Lambert law (supplementary material section A2), which relates the absorption rate to measurable quantities and constants, yielding 
\begin{widetext}
\begin{equation}\label{eq:DipoleAbs}
    D(\nu) = \left( \frac{3 \, \ln({10}) \, c_0 \, \varepsilon_0 \, h}{2 \, \pi^2 \, N_A } \right) \, \dd\nu  \, \left( \frac{1}{l \, \mathcal{C}} \right) \, \underbrace{\log_{10}\left(\frac{\mathcal{E}_{\mathrm{e},\nu,0}(\nu)}{\mathcal{E}_{\mathrm{e},\nu,\mathrm{t}}(\nu)}\right)}_{= A(\nu)} \, \frac{n(\nu)}{\nu}
\end{equation}
\end{widetext}
The frequency interval $\dd \nu$ can be considered as the bandwidth of the light source of the absorption spectrometer.\footnote{For the derivation of Eq.\,(\ref{eq:DipoleAbs}), we have assumed that the transitions within the frequency interval $\dd\nu$ possess a constant molar absorption coefficient or absorbance within $\dd\nu$.} Since most samples will feature broad absorption spectra at room temperature with linewidths $\gg \dd \nu$, the total dipole strength is obtained by integration over the absorption band.\cite{Parson2015}

As outlined earlier, spectrometers usually measure as a function of wavelength, and we therefore have to insert the measured wavelength-dependent quantities. Since only the ratio of measured quantities appears here, the frequency-to-wavelength conversion factors using the Jacobian transformation in Eq.\,(\ref{eq:JacobianConversion}) cancels out. Hence, in Eq.\,(\ref{eq:DipoleAbs}) only the wavelength dependence of the absorbance $A(\lambda)$ has to be exchanged to a frequency-dependence. This change will be indicated by $\lambda \rightarrow c_0 \, \nu^{-1}$ (see section \ref{sec:Jac}). Moreover, absolute values for the dipole strength are rarely required, since, often, only the shape of spectra and/or energy differences are relevant. The first bracket in Eq.\,(\ref{eq:DipoleAbs}), containing only constants, can then be dropped, and we arrive finally at
\begin{equation}
D(\nu)
   \propto  \dd\nu \, \frac{1}{l \, \mathcal{C}} \, A(c_0 \, \nu^{-1}) \, \left[\frac{n(\nu)}{\nu}\right]. \label{eq:AbsorptionFinalProp}
\end{equation}
Eq.\,(\ref{eq:AbsorptionFinalProp}) shows that the dipole strength is proportional to the absorbance, to the refractive index, and inversely proportional to the excitation frequency. The latter two quantities are highlighted between square brackets. These are the (often ignored) frequency-dependent correction factors that have to be applied when converting absorption data from wavelength to frequency dependence to obtain the dipole strength of the molecules, which is the relevant quantity describing intrinsic quantum-mechanical properties of molecules (or absorbers in general). This correction changes the shape of spectra, in particular, if absorption occurs over a wide spectral range (see section \ref{sec:Relevance}). It is important to note that the $\nu^{-1}$-scaling does not result from the wavelength-to-frequency conversion, but it results from the fact that the energy of a molecule increases by $h\nu$ upon absorption of a photon. As frequency is proportional to photon energy and wavenumber, this re-scaling in Eq.\,(\ref{eq:AbsorptionFinalProp}) also holds in those latter units.

As final words of caution, we emphasise that the full expression in Eq.\,(\ref{eq:DipoleAbs}) is derived and written using \textit{SI base units} for all quantities. Unfortunately, many textbooks use cgs units to describe the electromagnetic field, which results in a ``missing'' factor of $4\pi\epsilon_0$, and does not yield correct units (and wrong absolute values) if all the rest is inserted in SI units.\cite{Parson2015} Moreover, in the spectroscopy community, the absorption coefficient $\varepsilon$ is typically reported in L\,mol\textsuperscript{-1}\,cm\textsuperscript{-1}, rather than the SI standard m\textsuperscript{2}\,mol\textsuperscript{-1}. The reason for this is that more practical units are used: The path length $l$ is more practical in cm, the concentration $\mathcal{C}$ is traditionally reported in mol\,L\textsuperscript{-1}, the area $\mathcal{A}$ is in cm\textsuperscript{2}, and the spectral irradiance $\mathcal{E}$ is in W\,cm\textsuperscript{-2}\,Hz\textsuperscript{-1}. When using these ``spectroscopic units'' (instead of SI base units), the numerator of Eq.\,(\ref{eq:DipoleAbs}) must be multiplied by a factor 1000 to yield the correct value for $D(\nu)$.\footnote{Note that ``simply'' converting units of $\varepsilon(\nu)$ in Eq.\,(\ref{eq:DipoleAbs}) gives a factor of 10 in the denominator. In the derivation of this expression, however, the area on the sample container illuminated by the excitation beam must be considered as well, see supplementary material section A2. In ``spectroscopic units'' this area is in cm\textsuperscript{2}, which yields an additional factor of 10000 in the numerator for the conversion to SI base units.}


\subsection{Fluorescence}
\label{sec:Fl}
Once a molecule is in the excited state after an absorption process, it can relax back to the ground state via stimulated emission and spontaneous emission (fluorescence). Non-radiative processes, for instance, emission of phonons, although very common, are ignored here for conciseness. To derive an expression relating the dipole strength to the fluorescence rate $k_{\mathrm{fl}}$ per molecule, we start with the absorption rate in Eq.\,(\ref{eq:UprateIrradiance}). We rewrite this equation by expressing the spectral irradiance of the exciting light field using its spectral energy density $\rho_{\mathrm{e},\nu}(\nu)$, see supplementary material section B1, and obtain
\begin{equation}
    k_{\mathrm{abs}}(\nu) = \frac{2 \, \pi^2}{3 \, n(\nu)^{2}  \, \varepsilon_0 \, h^{2}} \, D(\nu) \, \rho_{\mathrm{e},\nu}(\nu) \equiv \mathbb{B}(\nu) \, \rho_{\mathrm{e},\nu}(\nu). 
    \label{eq:uprateEinstein}
\end{equation}
The last step defines the Einstein $\mathbb{B}$-coefficient for absorption [and stimulated emission, in units of s\textsuperscript{-1}\,(J\,m\textsuperscript{-3}\,Hz\textsuperscript{-1})\textsuperscript{-1}, see supplementary material section B2]. Eq.\,(\ref{eq:uprateEinstein}) links the $\mathbb{B}$ coefficient with the dipole strength $D(\nu)$. Exploiting the identity of the fluorescence rate and the Einstein $\mathbb{A}$ coefficient, as well as the relation between both Einstein coefficients [see Eq.\,(B10)], we get
\begin{eqnarray}\label{eq:Kfl}
k_{\mathrm{fl}}(\nu) = \mathbb{A}(\nu) &=& \frac{8 \, \pi \, h \, n(\nu)^3 \, \nu^3}{{c_0}^3} \, \mathbb{B}(\nu) \nonumber\\ 
&=& \frac{16 \, \pi^3 \, n(\nu) \, \nu^3}{3 \, h \, \varepsilon_0 \, {c_0}^3} \, D(\nu).
\end{eqnarray}
Hence, we find the dipole strength to be
\begin{equation}
    D(\nu) = \left( \frac{3 \, {c_0}^{3} \, \varepsilon_0 \, h}{16 \, \pi^3} \right ) \, k_{\mathrm{fl}}(\nu) \, \frac{1}{n(\nu) \, \nu^{3}}.
    \label{eq:ADRelation}
\end{equation}
Eq.\,(\ref{eq:ADRelation}) shows that the fluorescence rate and the dipole strength at the emission frequency $\nu$ are directly proportional. Since the dipole strength was derived from absorption, this implies that strong absorbers are strong emitters.\footnote{There are, of course, exceptions to this rule since we ignored non-radiative processes. But in general this holds true, for instance, for laser dyes, fluorescent labels in life sciences, etc.\cite{Lakowicz2006}} Another important aspect is the inverse cubic dependence on the emission frequency, which comes from the number of modes per unit volume. At high frequencies, more electromagnetic field modes per frequency interval are available for spontaneous emission. Finally, we note that the exponent of the refractive index $n(\nu)$ is different from that of the frequency and the speed of light. To derive Eq.\,(\ref{eq:ADRelation}) the spectral energy density has to be evaluated in the medium surrounding the molecules, which cancels out a factor of $n(\nu)^2$.

The final step to obtain a tractable expression to determine $k_{\mathrm{fl}}$ is to connect the fluorescence rate in Eq.\,(\ref{eq:ADRelation}) with wavelength-dependent quantities measured in spectrometers. Fluorescence of a sample with many molecules is isotropic in space, but only photons emitted within an elementary solid angle $\dd \Omega_{\mathrm{det}}$ in the direction of the detector will contribute to the signal. This signal should therefore be expressed as a function of directional radiometric quantities, e.g. the spectral radiant intensity $I_{\mathrm{e},\lambda,\Omega}(\lambda)$ (in W\,sr$^{-1}$\,m$^{-1}$, see section \ref{sec:Measure}). After integration over $\dd\Omega_{\mathrm{det}}$, this yields the spectral radiant power received by the detector $P_{\mathrm{e},\lambda,\mathrm{d}}(\lambda)$. Since $\dd \Omega_{\mathrm{det}}$ is in practice determined by the entrance slit of the monochromator, the spectrometer's apertures, and the active area of the detector, we assume it to be constant, independent of the fluorescence wavelength. Overall, a constant (but unknown) fraction of all emitted photons will thus contribute to the detected radiant power, which can be written using Eq.\,(\ref{eq:QE_responsivity}) as $P_{\mathrm{e,d}}(\lambda) = q_{\mathrm{p,d}}(\lambda) \, h \, c_0 / \lambda.$ The detected photon flux $q_{\mathrm{p,d}}$ is proportional to the product of the number of molecules in the excited state $N_\mathrm{e}$ and the fluorescence rate $k_{\mathrm{fl}}(\lambda)$ per molecule. $N_\mathrm{e}$ is in practice also unknown, but it is a constant value for given excitation conditions. Hence,
\begin{eqnarray}\label{eq:RadiantIntensity}
k_{\mathrm{fl}}(\lambda) & \propto & q_{\mathrm{p,d}}(\lambda) \\ 
& \propto & P_{\mathrm{e,d}}(\lambda) \, \frac{\lambda}{h\cdot c_0} \nonumber 
\end{eqnarray}
The detected fluorescence signal as a function of wavelength $S_\mathrm{d}(\lambda)$ (here the photon flux or the radiant power) must be converted to the frequency scale via the Jacobian transformation [Eq.\,(\ref{eq:JacobianConversion})]. We obtain for the frequency-dependent dipole strength [dropping all constants coming from Eqs.\,(\ref{eq:JacobianConversion}), (\ref{eq:ADRelation}), and (\ref{eq:RadiantIntensity})]
\begin{equation}\label{eq:FlFinalProp}
    D(\nu) \propto
    \begin{cases}
    S_\mathrm{d}(c_{0} \, \nu^{-1}) \, \left[ \frac{1}{n(\nu) \, \nu^{5}} \right] & \text{when $S_\mathrm{d}$ = $q_{\mathrm{p,d}}$},\\
     S_\mathrm{d}(c_{0} \, \nu^{-1}) \, \left[ \frac{1}{n(\nu) \, \nu^{6}} \right]& \text{when $S_\mathrm{d}$ = $P_{\mathrm{e,d}}$.}
    \end{cases}
\end{equation} 

For simplicity we left out the correction of instrumental effects via the correction function $\Theta(\lambda)$ [Eq.\,(\ref{eq:CorrectedSpectrum})], as well as the detector's quantum efficiency, respectively, the spectral responsivity. If not not taken into account by the instrument, this correction must be done before applying Eqs.\,(\ref{eq:FlFinalProp}). The full expressions can be found in the supplementary material Eq. (B11) and (B12). We emphasize that different powers of frequency appear in Eq.\,(\ref{eq:FlFinalProp}) depending on whether a photon flux or a radiant power is measured. It is therefore important to carefully check which (wavelength-dependent) signal is actually measured by the instrument.


\subsection{Fluorescence excitation}
\label{sec:Ex}
In fluorescence excitation spectroscopy the fluorescence signal is measured as a function of the excitation wavelength to retrieve the absorbance of weakly absorbing samples (see section \ref{sec:SpecTec}). To connect the measured fluorescence signal to the absorption rate, we exploit the fluorescence quantum yield $\Phi = \frac{k_{\mathrm{fl}}}{k_{\mathrm{abs}}}$ introduced earlier. Using further the relation between the absorption rate, the dipole strength, and the spectral irradiance of the excitation light in Eq.\,(\ref{eq:UprateIrradiance}), we obtain the expression for the dipole strength at excitation frequency $\nu_{\mathrm{ex}}$ as a function of the fluorescence rate measured:
\begin{equation}
    D(\nu_{\mathrm{ex}}) = \frac{3 \, \epsilon_0 \, c_0 \, h^2}{2 \, \pi^2 \, \Phi} \, \frac{k_{\mathrm{fl}}(\nu_{\mathrm{ex}})}{\mathcal{E}_{\mathrm{e},\nu,0}(\nu_{\mathrm{ex}})} \, n(\nu_{\mathrm{ex}}). 
    \label{eq:DipoleExcitation}
\end{equation}

Since the fluorescence rate $k_{\mathrm{fl}}$ is measured at a constant fluorescence frequency, the instrumental corrections in Eq.\,(\ref{eq:CorrectedSpectrum}) are all constant. For the same reason, $k_{\mathrm{fl}}$ is directly proportional to photon flux and radiant power, and the wavelength in Eq.\,(\ref{eq:RadiantIntensity}) is constant. We can thus express $k_{\mathrm{fl}}$ in terms of the detected fluorescence signal $S_{\mathrm{d,fl}}$, measured as a function of the excitation frequency or wavelength. Finally, we convert the incident spectral irradiance $\mathcal{E}_{\mathrm{e},\nu,0}(\nu_{\mathrm{ex}})$ from the frequency to the wavelength axis. These measurements have to be performed at low concentrations, so that we can safely assume that the spectral irradiance is constant across the sample at each excitation wavelength. However, since the spectral irradiance of the excitation light source is usually a (strong) function of this excitation frequency, this quantity has to be measured separately (see Fig.\,\ref{fig:ExcitationSpectrum}c) and can thus not be ignored. Assuming that the illuminated area on the detector is identical for all excitation frequencies, the spectral irradiance of the light source is proportional to the spectral radiant power $P_{\mathrm{e},\nu,0}(\nu_{\mathrm{ex}})$ or to the product of spectral photon flux and excitation frequency $q_{\mathrm{p},\nu,0}(\nu_{\mathrm{ex}}) \, \nu_{\mathrm{ex}}$. Like before, as the tuning of the excitation source occurs in equidistant wavelength steps, we need a Jacobian transformation. This modifies Eq.\,(\ref{eq:DipoleExcitation}) into
\begin{equation}\label{eq:ExFinalProp}
    D(\nu_{\mathrm{ex}}) \propto
    \begin{cases}
    \left(\frac{S_{\mathrm{d,fl}}(c_{0}/\nu_{\mathrm{ex}})}{q_{\mathrm{p},\nu,0}(c_{0}/\nu_{\mathrm{ex}})} \right)  \, \left[ n(\nu_{\mathrm{ex}})^2 \, \nu_{\mathrm{ex}}\right] & \text{for $q_{\mathrm{p,d}}$},\\
      \left(\frac{S_{\mathrm{d,fl}}(c_{0}/\nu_{\mathrm{ex}})}{P_{\mathrm{e},\nu,0}(c_{0}/\nu_{\mathrm{ex}})} \right)  \, \left[ n(\nu_{\mathrm{ex}})^2 \, {\nu_{\mathrm{ex}}}^{2} \right] & \text{for $P_{\mathrm{e,d}}$.}
    \end{cases}
\end{equation}
The additional factor $n(\nu_{\mathrm{ex}})$ compared to Eq.\,(\ref{eq:DipoleExcitation}) results from the spectral irradiance of the excitation light in the medium surrounding the molecules. Thus, in the Jacobian transformation in Eq.\,(\ref{eq:JacobianConversion}) the speed of light in vacuum must be substituted by the speed of light in the medium. Even though we are dealing with a ratio of signals, one still needs to perform corrections based on the optical elements between the sample and the detector, and on the detector themselves [Eq.\,(\ref{eq:CorrectedSpectrum})]. This does not arise from the fluorescence signal, as any correction here is constant due to $\nu_{\mathrm{fl}}$ staying constant, but rather from the reference signal of the light source, which has to be corrected for [and which we left out for simplicity in Eq.\,(\ref{eq:ExFinalProp})]. The full expressions can be found in the supplementary material section C.\\


\section{Worked example} 
\label{sec:Relevance}

In this section, we demonstrate the effect of converting measured (raw) spectra into spectra displaying a signal proportional to the dipole strength using the 1,4-Bis(5-phenyl-2-oxazolyl)benzene (POPOP) molecule as an example (for a second example see supplementary material section D). The raw absorption and fluorescence spectra as a function of wavelength were taken from an online database.\cite{Taniguchi2017} The spectra of POPOP dissolved in cyclohexane were measured with commercial spectrometers, i.e., instrumental effects [Eq.\,(\ref{eq:CorrectedSpectrum})] are already corrected using stored functions.

\begin{figure}[h!]
\centering
\includegraphics[width=8.5cm]{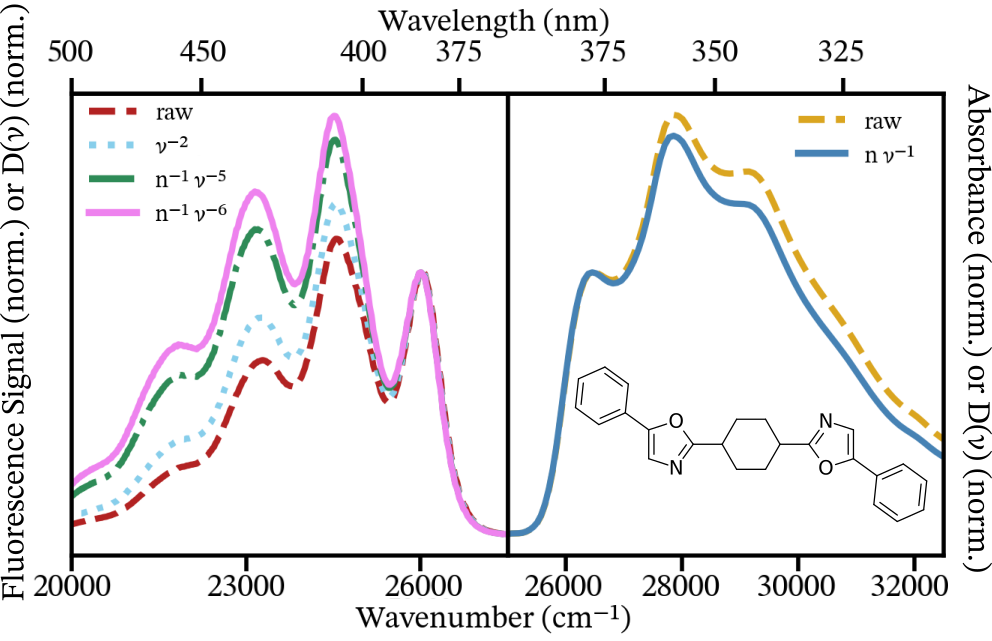}
\caption{(Color online) Fluorescence (left) and absorption (right) spectra of the POPOP molecule (see right inset for the chemical structure) with successive application of all corrections discussed in this work. For the fully corrected spectra (solid), the y-axis represents a normalized dipole strength. The raw spectra (dashed) have been taken from Ref. \onlinecite{Taniguchi2017}}
\label{fig:POPOP_Full}
\end{figure}

\begin{figure*}[h!]
\centering
\includegraphics[width=15cm]{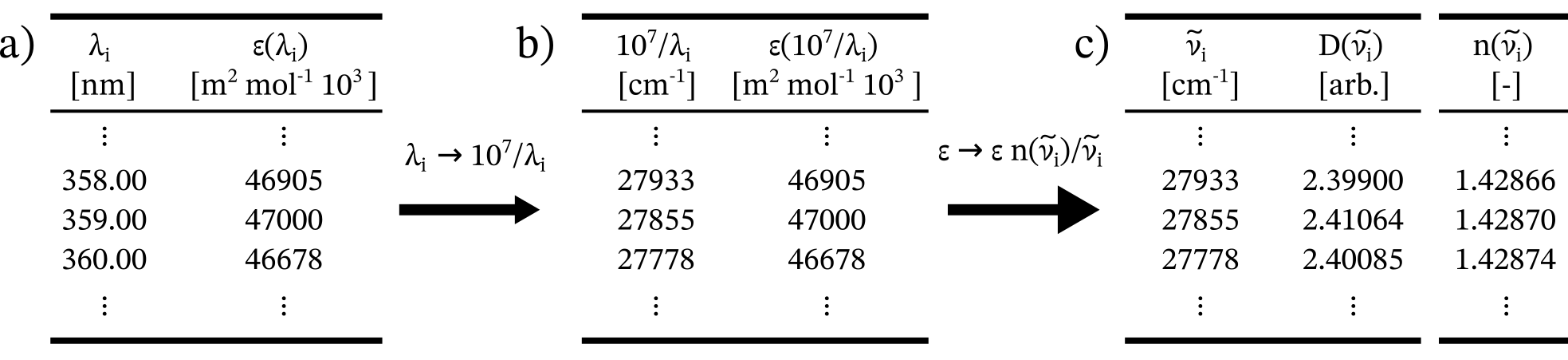}
\caption{\textit{Step-by-step conversion of the data file for the absorption spectrum of the POPOP molecule. In c) we have also added a column with the values of the refractive index $n(\Tilde{\nu_i})$ of the solvent.\cite{Kozma2005}} For details see text.}
\label{fig:ConvertTable}
\end{figure*}

The corrections discussed above are illustrated step-by-step in Fig.\,\ref{fig:POPOP_Full} for the absorption (right) and fluorescence spectra (left) of the POPOP molecule. Furthermore, we show a small part of the absorption spectrum data file in Fig.\,\ref{fig:ConvertTable} to illustrate the numerical conversion process. The absorption spectrum shown as a dashed yellow line is the ``raw'' (as measured) spectrum where we have only changed the original wavelength scale (in nm) to a wavenumber scale (in cm$^{-1}$). In the original data file (Fig.\,\ref{fig:ConvertTable}a), the molar absorption coefficient $\varepsilon(\lambda_\mathrm{i})$ was tabulated for each wavelength $\lambda_\mathrm{i}$ in steps of 1\,nm. Hence, this change of the horizontal axis only required to replace the wavelength data column by a wavenumber column according to $\Tilde{\nu}_\mathrm{i} = 10^7 / \lambda_\mathrm{i}$ (see Fig.\,\ref{fig:ConvertTable}b --- using nm as unit for $\lambda$, this directly yields cm$^{-1}$ for $\Tilde{\nu}$). The molar absorption coefficient values $\varepsilon(10^7/\lambda_\mathrm{i})$ remain unchanged, since it represents the ratio between two measured spectral quantities (typically spectral radiant power or spectral irradiance). The Jacobian transformation is therefore not required. To obtain a quantity proportional
to the dipole strength, the factor of $n(\nu)\,\nu^{-1}$ has to be applied [Eq.\,(\ref{eq:AbsorptionFinalProp})]. This factor changes the spectrum to that shown as solid blue in Fig.\,\ref{fig:POPOP_Full} and the data column to that in Fig.\,\ref{fig:ConvertTable}c. Since we are only interested in the relative dipole strengths for the different absorption peaks, we finally normalize both the raw and the corrected absorption spectra in Fig.\,\ref{fig:POPOP_Full} to the lowest-energy peak at 26455\,cm$^{-1}$ (corresponding to 378\,nm). This normalisation illustrates that the relative amplitudes of the absorption peaks change when converted to the dipole strength. 

For fluorescence (Fig.\,\ref{fig:POPOP_Full}, left), the ``raw'' spectrum was first converted to the wavenumber scale. Then the measured fluorescence signal was subjected to the necessary Jacobian transformation (blue dotted), since the fluorescence detector outputs a signal proportional to a spectral quantity (e.g. spectral radiant power or spectral irradiance). Next, the correction for the mode density of the electromagnetic field (correction factor $n^{-1}\nu^{-3}$) is applied (green dash-dotted). Finally, since the detector outputs a spectral quantity, which is proportional to the detected photon energy, an additional factor of $\Tilde{\nu}^{-1}$ is required (pink, solid) to obtain the final fluorescence signal that is proportional to the dipole strength. The spectra are all normalized to the highest-energy peak to highlight the strong effect of the conversions on the spectral shape.

From Fig.\,\ref{fig:POPOP_Full} it can be clearly seen that the required conversions for both fluorescence and absorption spectra have some impact on the spectrum. Here, we especially want to highlight the change in ratio between the highest energy peak (at 26000 \,cm$^{-1}$) and its neighboring peak (at 24500\,cm$^{-1}$) in the fluorescence spectra. If no correction is applied to the spectrum, the ratio between the amplitudes of the two peaks would be 1:1.1. By applying the Jacobian transformation, this ratio increases to 1:1.3. Applying the full conversion for measured photon flux and radiant power, we obtain a peak ratio of 1:1.5 and 1.6, respectively. A more in-depth analysis regarding the underlying quantum physics can be found in the supplementary material section A4. 

We also want to stress that we labelled the y-axis of the fluorescence spectra as ``Fluorescence Signal''. Often ``Intensity'' is used in literature, which is, to some extent, justified if raw data are shown, since the radiant power, the radiant intensity, and the irradiance are all related to each other by factors that are constant for a given experimental setup (e.g. a solid angle, illuminated area on the detector, etc., see section \ref{sec:Measure}).\footnote{Note that only ''radiant intensity'' with units W sr$^{-1}$ is a SI quantity, whereas in the spectroscopy community, irradiance with units W\,cm$^{-2}$ is sometimes called intensity, which is incorrect.} For corrected spectra a quantity proportional to dipole strength should not be labelled ``intensity'', ``power'' or ``irradiance'', since it is proportional to the detected photon frequency (energy, wavenumber), which has to be corrected for. To have a intuitive label we opted for ``signal'' here.


\section{Concluding Remarks}
In this paper, we have outlined the corrections that have to be applied to wavelength-dependent (optical) spectra in order for them to reflect frequency- (photon-energy-) dependent intrinsic quantum-mechanical properties of molecules. For conciseness, our description has been simplified in certain aspects. For instance, we only introduced measurements using a monochromator to obtain spectra. In particular, in infrared spectroscopy Fourier transform-based techniques are commonly employed, which come with their own challenges.\cite{Griffiths2007} Moreover, the description of absorbers/emitters as two-level systems is in most cases far from realistic. We have, however, pointed out how our description can be extended to the more complex electronic structure of real systems (here: organic molecules). Furthermore, we have restricted the description to narrow-band excitation of broad spectra, which is often sufficient at room temperature. A modification to describe excitation with e.g. broad-band laser pulses is of course possible.\cite{Parson2015} Finally, we followed Einstein's derivation of the relationships between absorption, stimulated emission and fluorescence rates and the corresponding Einstein coefficients. However, since Einstein assumed infinitely narrow lines, but a finite lifetime of the excited level of the absorbing/emitting system, this derivation is inconsistent with Heisenberg's time-energy uncertainty principle. More general approaches have been developed, including approaches that use realistic linewidths of excited levels of real systems.\cite{Ryu2024, McCumber1964, Fowler1962} Those approaches change absolute values, but still yield the frequency-scaling factors $\nu^{-1}$ for absorption and $\nu^{-3}$ for fluorescence (without Jacobian transformation) that have to be applied prior to quantitative analysis, as we have outlined here. We hope that our overview will be helpful to instructors during the training of students, as well as to more experienced users and researchers as a reference. 

\section{Supplementary Material}
Please click on this link to access the supplementary material, which includes detailed derivations, detailed quantum mechanical properties of molecules, and a second worked example. Print readers can see the supplementary material at [DOI to be inserted by AIPP].

\section{Author Declaration}
The authors have no conflicts to disclose.


\begin{acknowledgments}

We gratefully acknowledge financial support by Nederlandse Organisatie voor Wetenschappelijk Onderzoek (NWO) through grant OCENW.KLEIN.500. We are also grateful to Prof. R.C. Chiechi (NCSU) for providing the N2200 sample.

\end{acknowledgments}


\bibliographystyle{ieeetr} 
\bibliography{MainText} 


\end{document}